\documentclass[usenatbib]{mn2e}
\usepackage{graphicx}
\title[New approach for modeling of transiting exoplanets]
{New approach for modeling of transiting exoplanets for arbitrary limb-darkening law}

\author[D. Kjurkchieva, D. Dimitrov, A. Vladev, V. Yotov]{Diana Kjurkchieva$^{1}$
\thanks{E-mail: d.kyurkchieva@shu-bg.net; dinko@astro.bas.bg; avladev@gmail.com; vergil.yotov@gmail.com}
Dinko Dimitrov$^{2}$, Anatoli Vladev$^{1}$ and Vergil Yotov$^{3}$\\
$^{1}$Department of Physics, Shumen University, 9700 Shumen, Bulgaria\\
$^{2}$Institute of Astronomy and NAO, Bulgarian Academy of Sciences,
Tsarigradsko shossee 72, 1784 Sofia, Bulgaria\\
$^{3}$School of Physics and Astronomy, The University of
Edinburgh, Edinburgh, EH9 3JZ, United Kingdom\\}

\begin{document}

\date{Accepted 2013 March 10. Received 2013 March 10; in original form 2012 October 15}

\pagerange{\pageref{firstpage}--\pageref{lastpage}} \pubyear{2013}

\maketitle

\label{firstpage}

\begin{abstract}
We present a new solution of the direct problem of planet transits based on transformation of double integrals to single ones. On the
basis of our direct problem solution we created the code \textsc{TAC-maker} for rapid and interactive calculation of synthetic planet 
transits by numerical computations of the integrals. The validation of our approach was made by comparison with the results of the wide-spread 
Mandel \& Agol (2002) method for the cases of linear, quadratic and squared root limb-darkening laws and various combinations of model 
parameters. For the first time our approach allows the use of arbitrary limb-darkening law of the host star. This advantage together with 
the practically arbitrary precision of the calculations make the code a valuable tool that faces the challenges of the continuously 
increasing photometric precision of the ground-based and space observations.
\end{abstract}


\section{Introduction}\label{sec:intro}

The determination of geometric parameters of extrasolar planets has an essential role in inferring their densities and hence their compositions, 
masses and ages. This information leads to refinements of the models of planetary systems and yields important constraints on planet formation
\citep{cody02, hubb01,seager03}.

During the last several years there is a sharp rise in the detections of transiting extra-solar planets (TEP) mainly by the wide-field 
photometric variability surveys: (i) ground-based observations as SuperWASP \citep{poll06}, HATNet \citep{bakos04}, OGLE-III \citep{udal02}, 
TrES \citep{alon04}, etc.; (ii) space missions as \textit{CoRoT} \citep{bagl06} and \textit{Kepler} \citep{bor10a}.

The \textit{Kepler} mission produced a real bump of the number of exoplanet candidates, dozens of them yet confirmed 
\citep[][etc.]{koch10, bor10b, dunham10, latham10}. The space-based missions, especially \textit{Kepler}, have the photometric ability 
to detect even transiting terrestrial-size planets \citep{stef12}.

The recent increasing number of the planet-candidate discoveries and the increasing precision of the observations allow to investigate 
fine effects such as:

\begin{description}

\item[(a)] rotational and orbital synchronization and alignment \citep{winn11, hebr10}; 
\item[(b)] zonal flows and violent atmospheric dynamics due to large temperature contrast between day-sides and night-sides of the planets 
\citep{burr10};
\item[(c)] departures from sphericity of the planets and spin precession \citep{carter10}; 
\item[(d)] rings and satellites \citep{hui02, arnold06}; 
\item[(e)] stellar spots \citep{rabus09, ditt09, hebr10}; 
\item[(f)] dependence of the derived planetary radius from the limb-darkening coefficients \citep{kipp10};
\item[(g)] irradiation by the parent star \citep{seager00b}, planet transmission spectra \citep{seager00a}, atmospheric lensing due to 
atmospheric refraction \citep{hui02}, Rayleigh scattering, cloud scattering, refraction and molecular absorption of starlight in the 
planet atmosphere \citep{hubb01}, etc.
\end{description}

The study of such fine effects requires very precise methods for determination of parameters of the planetary systems from the observational data.

Recently, we established that the synthetic light curves generated by the widely-used codes for planet transits deviated from the expected 
smooth shape. This motivated us to search for a new approach for the direct problem solution of the planet transits. We managed to realize 
this idea successfully for the case of orbital inclination $i=90\degr$ and linear limb-darkening law \citep{kjurk12}. This paper presents 
continuation of the new approach for arbitrary orbital inclinations and arbitrary limb-darkening laws.

\section{Methods for solution of the planet transit problem}

\subsection{Previous approaches}

The global parameters of the configurations of TEPs are different from those of eclipsing binary systems. The main geometric difference is 
that the radii of the two components of TEPs are very different, while those of EBs are comparable. As a result, almost all models based on 
numerical integration over the stellar surfaces of the components give numerical errors, especially around the transit center. The overcoming 
of this problem required specific approaches and models for the study of TEPs.

The first solution of the direct problem of the planet transits is that of \citet{mandel02}. They derived analytical formulae describing the 
light decreasing due to covering of stellar disk by a dark (opaque) planet in cases of quadratic and nonlinear limb-darkening laws. 
The formulae of \citet[][further M\&A solution]{mandel02} \defcitealias{mandel02}{M\&A} contain several types of special functions 
(beta function, Appel's hypergeometric function and Gauss hypergeometric function, complete elliptic integral of the third kind). 
To generate synthetic transits \citet{mandel02} created \textsc{IDL} and \textsc{Fortran} codes \textsc{occultsmall} (for small planet), 
\textsc{occultquad} (for quadratic limb-darkening law) and \textsc{occultnl} (for nonlinear limb-darkening laws) that are based on numerical 
calculations of the special function values. These codes are widely used by many investigators for analysis of observed transits and improved 
later by different authors.

In the meantime \citet{seager03} obtained analytical solution for the particular case of total transit and uniform stellar disk (i.e. neglecting 
the limb-darkening effect), \citet{knut07} made calculation of the secondary planet eclipses while \citet{kipp08,kipp10b} studied the problem 
of eccentric orbits.

The second direct problem solution for the planet transits was made by \citet{gim06}. He derived analytical formulae for the computation the 
light curves of planet' transits for arbitrary limb-darkening laws. This approach is similar to that of the Kopal's $\alpha_{\rm{n}}$ functions 
and the derived formulae contain different special functions (elliptic integrals of the first, second and third order).

Most of the codes for inverse problem solutions of planet transits are based on the \citetalias{mandel02} solution. For instance, the recent 
packages \textsc{TAP} and \textsc{autoKep}\footnote{http://ifa.hawaii.edu/users/zgazak/IfA/TAP.html} \citep{gazak12}. 
\citet{pal10} estimated the fit quality of the inverse problem corresponding to the \citetalias{mandel02} solution. It should be noted that 
the inverse problem solution of the planet transits is not a trivial task. The known codes for stellar eclipses are not applicable for the
analysis of most planet transits due to the non-effective convergence of the differential corrections in cases of observational precisions 
poorer than 1/10 the depth of planet transit. \textsc{EBOP} \citep{etzel75,etzel81,popper81} is the only model for EBs, which heavily 
modified version \textsc{JKTEBOP} \citep{south08,south04a,south04b} can be applied successfully to TEPs.

Recently, solutions of the whole inverse problem based on simultaneous modeling of photometric and spectral data of exoplanets performed 
using the Markov-chain Monte Carlo (MCMC) code were proposed \citep[][etc.]{coll07,poll08}. But their subroutines for fitting of the transits 
also use the \citetalias{mandel02} solution for quadratic limb-darkening law. For instance, the subroutine \textsc{exofast\_occultquad} of the
package \textsc{exofast} \citep{eastman13} is a new improved version of \textsc{occultquad}.

An opportunity to generate synthetic transit light curve as well as to search for fit to own observational data is provided from the website 
Exoplanet Transit Database\footnote{http://var2.astro.cz/ETD/protocol.php} (ETD). It is based on the \textsc{occultsmall} routine of the 
\citetalias{mandel02} solution that uses the simplification of the planet trajectory as a straight line over the stellar disk \citep{podd10}.

\subsection{The new solution of the direct problem: how to use the symmetry of the problem to reduce the double integral to a single one}

Let's consider configuration from a spherical planet with radius $R_{\rm{p}}$ orbiting a spherical star with radius $R_{\rm{s}}$ on circle orbit 
with radius $a$, period $P$ and initial epoch $T_{0}$. Let's the line-of-sight is inclined at an angle $i$ to the orbital plane of the planet.

Usually, it is assumed that the limb-darkening of the main-sequence stars may be represented by the linear function

\begin{equation}\label{eq01}
I(\mu)=I_{0}[1 - u(1 - \mu)]
\end{equation}

\noindent where $I_{0}$ is the light intensity at the center of the stellar disk depending on the stellar temperature, $\mu = \cos
\theta$ and $\theta$ is the angle between the normal to the current point of the stellar surface and the line of sight.

\citet{claret00} found that the more accurate limb-darkening functions are the quadratic law \citep{kopal50}

\begin{equation}\label{eq02}
I(\mu)=I_{0} \left[ 1 - u_{1} (1 - \mu) - u_{2}( 1-\mu )^2 )
\right]
\end{equation}

\noindent and ``nonlinear'' law

\begin{equation}\label{eq03}
I(\mu)=I_{0} \left[ 1 - \sum_{j=1}^{4} u_{\rm{j}} ( 1- \mu
^{\rm{j/2}}) \right]
\end{equation}

\noindent which is a Taylor series in $\mu$ to fourth order in 1/2.

Square-root law is proposed by \citet{diaz92}

\begin{equation}\label{eq04}
I(\mu)=I_{0} \left[ 1 - u_{1} (1 - \mu) - u_{2}( 1-\sqrt {\mu} )
\right]
\end{equation}

\noindent while \citet{kling70} proposed logarithmic law for early stars

\begin{equation}\label{eq05}
I(\mu)=I_{0} \left[ 1 - u_{1} (1 - \mu) - u_{2} \mu \ln {\mu}
\right].
\end{equation}

Our solution can be applied for arbitrary limb-darkening law

\begin{equation}\label{eq06}
I(\mu)=I_{0} f(u_{\rm{j}}, \mu)
\end{equation}

\noindent where $f(u_{\rm{j}}, \mu )$ is an arbitrary function of $\mu$.

The possibility to use an arbitrary limb-darkening law is one of the main advantages of our approach.

The luminosity of the planetary system at phase $\varphi$ out-of-transit is

\begin{equation}\label{eq07}
L(\varphi)= L_{\rm{s}}+L_{\rm{p}}(\varphi)
\end{equation}

\noindent where $L_{\rm{s}}$ is the stellar luminosity and $L_{\rm{p}}(\varphi)$ is the planet luminosity. The phase $\varphi$ is calculated by 
the period $P$ and the initial epoch $HJD(\rm{min})$.

The planet luminosity $L_{\rm{p}}(\varphi)$ is variable out of the eclipse (because more of the day-side of the planet is visible after the 
occultation while more of the night-side of the planet is visible before the transit). Due to the relatively small size and low temperature 
of the planet, it can be assumed that its disk is uniform and its luminosity does not change during the transit, i.e.

\begin{equation}\label{eq08}
L_{\rm{p}}=\pi R_{\rm{p}}^2 I_{\rm{p}}
\end{equation}

\noindent where $I_{\rm{p}}$ depends on the planet temperature.

The luminosity during the transit is

\begin{equation}\label{eq10}
L(\varphi)= L_{\rm{s}}+L_{\rm{p}}-\tilde{J}(\varphi)
\end{equation}

\noindent where $\tilde{J}(\varphi)$ is the light decrease due to the covering of the star by the planet. It might be expressed in the form

\begin{equation}\label{eq11}
\tilde{J}(\varphi)=\int
\limits_{S_{\rm{oc}}(\varphi)}I_{0}f(u_{\rm{j}},\mu) ds
\end{equation}

\noindent where the integration is on the stellar area $S_{\rm{oc}}(\varphi)$ covered by the planet. Hence, the solution of the direct problem 
for the planet transit is reduced to a calculation of a surface integral. It is not a trivial task due to the nonuniform-illuminated stellar disk.

If we assume the out-of-transit flux to be $F_{\rm{out}}$=1, then its decreasing during the transit is described by the expression

\begin{equation}\label{eq12}
F(\varphi)= \frac {L_{\rm{s}}+L_{\rm{p}}-\tilde{J}(\varphi)}{L_{\rm{s}}+L_{\rm{p}}}
\end{equation}

For the next considerations we use coordinate system whose origin coincides with the stellar center. The axis $z$ is along the line-of-sight and 
the $xy$ plane coincides with the visible plane. We choose the axis $y$ to be along the projection of the normal to the orbit on the visible plane.

\begin{figure}
 \centering
 \includegraphics[width=0.99\columnwidth]{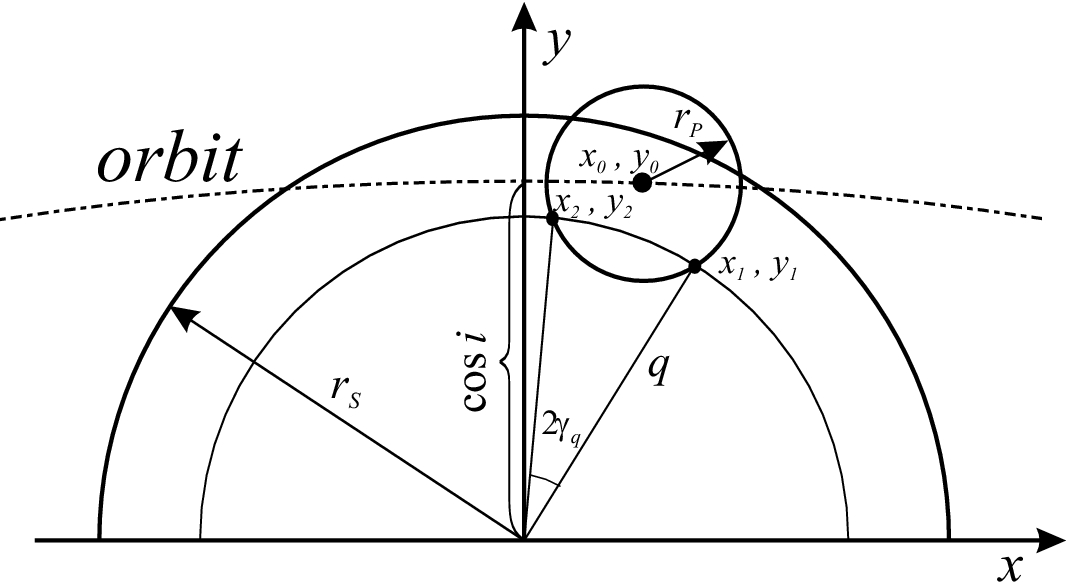}
 \caption{The visible plane: geometry of the partial transit (Case A.1)}
 \label{fig01}
\end{figure}

Taking into account that the stellar isolines of equal light intensity are concentric circles with radius $r$ we may calculate the light decrease 
as a sum (integral) of the contributions of differential uniformly-illuminated arcs with central angles 2$\gamma_{\rm{r}}$ and area 
$ds=2 \gamma_{\rm{r}} r dr $ (Fig. \ref{fig01}). In this way we transform the surface integral (\ref{eq11}) to a linear one

\begin{equation}\label{eq13}
\tilde{J}(\varphi)=\int \limits_{r_{\rm{min}}(\varphi)}^{r_{\rm{max}}(\varphi)} I_{0} f(u_{\rm{j}}, \mu) 2 \gamma_{\rm{r}}(\varphi) r dr
\end{equation}

\noindent where

\begin{equation}\label{eq14}
\mu = \cos \theta = \sqrt {1-\left( \frac{r}{R_{\rm{s}}} \right)^2}.
\end{equation}

The integrand of our main equation (\ref{eq13}) is different from that of the main equation (2) of the \citetalias{mandel02} solution. As a result, 
the methods of numerical calculation of these integrals are different.

The integration limits $r_{\rm{min}}(\varphi)$ and $r_{\rm{max}}(\varphi)$ are the extremal radii of the stellar isolines that are covered by the 
planet at orbital phase $\varphi$. These limits depend on the configuration parameters.

It is appropriate to assume the separation $a$ as a size unit and to work with dimensionless quantities: relative radius of the planet 
$r_{\rm{p}}=R_{\rm{p}}/a$; relative radius of the star $r_{\rm{s}}=R_{\rm{s}}/a$; relative radius of the stellar isoline $q=r/a$.

Then the equation for the light decrease during the transit can be rewritten in the following form:

\begin{equation}\label{eq15}
F(\varphi) = F_{\rm{out}} - \frac {J(\varphi)}{\pi kr_{\rm{p}}^2+L_{\rm{s}}/[I_{0} a^{2}]}
\end{equation}

\noindent where

\begin{equation}\label{eq16}
J(\varphi)= \int \limits_{q_{min}(\varphi)}^{q_{max}(\varphi)}
f(u_{\rm{j}},\mu) 2\gamma_{q}(\varphi) q dq
\end{equation}

\begin{equation}\label{eq17}
k=I_{\rm{p}}/I_{0}=[T_{\rm{p}}/T_{0}]^4
 \end{equation}

Assuming the center of the transit to be at phase 0.0 we consider only the phase interval (0, 0.5) because in the case of spherical star and planet 
the transit light curve in the range (-0.5, 0) is symmetric to that in the range (0, 0.5).

For a circular orbit the coordinates (in units $a$) of the planet center at phase $\varphi$ are

\begin{eqnarray}\label{eq18}
x_{0}(\varphi) & = & \sin (2\pi \varphi)\nonumber \\
y_{0}(\varphi) & = & \cos i \cos(2 \pi \varphi).
\end{eqnarray}

The coordinates $x_{1,2}(\varphi), y_{1,2}(\varphi)$ (in units $a$) of the intersection points of the stellar brightness isoline with radius $q$ 
and the planet limb are respectively (Fig. \ref{fig01}):

\begin{eqnarray}\label{eq19}
x_{1}(\varphi) & = & \frac {c^2(\varphi) x_{0}(\varphi)+y_{0}(\varphi) \sqrt {4b^2(\varphi) q^2-c^4(\varphi)}} {2b^2(\varphi)}\nonumber \\
x_{2}(\varphi) & = & \frac {c^2(\varphi) x_{0}(\varphi)-y_{0}(\varphi) \sqrt {4b^2(\varphi) q^2-c^4(\varphi)}} {2b^2(\varphi)}\nonumber \\
y_{1}(\varphi) & = & \frac {c^2(\varphi) y_{0}(\varphi)-x_{0}(\varphi) \sqrt {4b^2(\varphi) q^2-c^4(\varphi)}} {2b^2(\varphi)}\nonumber \\
y_{2}(\varphi) & = & \frac {c^2(\varphi)
y_{0}(\varphi)+x_{0}(\varphi) \sqrt {4b^2(\varphi)
q^2-c^4(\varphi)}} {2b^2(\varphi)}.
\end{eqnarray}

\noindent where we have introduced the designations

\begin{eqnarray}\label{eq20}
b^2 (\varphi)& = & x_{0}^{2}(\varphi)+y_{0}^{2}(\varphi)=1-\cos^2 (2\pi\varphi) \sin^2 i \nonumber \\
c^2 (\varphi)& = & q^2+b^2(\varphi)-r_{\rm{p}}^2.
\end{eqnarray}

Further we will derive the expressions for the angle $2\gamma_{\rm{q}}(\varphi)$ and the integration limits in equation (\ref{eq16}) for different 
combinations of geometric parameters and at different phases.

\subsubsection{Case A: Partial transits}

The transit is partial if the orbital inclination is into the
range $i_{1}\leq i \leq i_{2}$ where
\begin{eqnarray}\label{eq21}
i_{1} & = & \arccos (r_{\rm{s}}+ r_{\rm{p}}) \nonumber \\
i_{2} & = & \arccos (r_{\rm{s}}- r_{\rm{p}}) .
\end{eqnarray}

The phases of outer contacts star-planet are $-\varphi_{1}$ (beginning of the transit) and $+\varphi_{1}$ (end of the transit) where

\begin{equation}\label{eq22}
\varphi_{1}= \frac {1}{2 \pi} \arcsin \frac {\sqrt{(r_{\rm{s}}+r_{\rm{p}})^2 - \cos i^2}}{\sin i}.
\end{equation}

The partial transit occurs into the phase range [0,~$\varphi_{1}$]. For the sake of brevity we will not write further the dependence of
$x_{0}, y_{0}, x_{1}, x_{2}, y_{1}, y_{2}, b, c$ on $\varphi$.

\begin{figure}
 \centering
 \includegraphics[width=0.99\columnwidth]{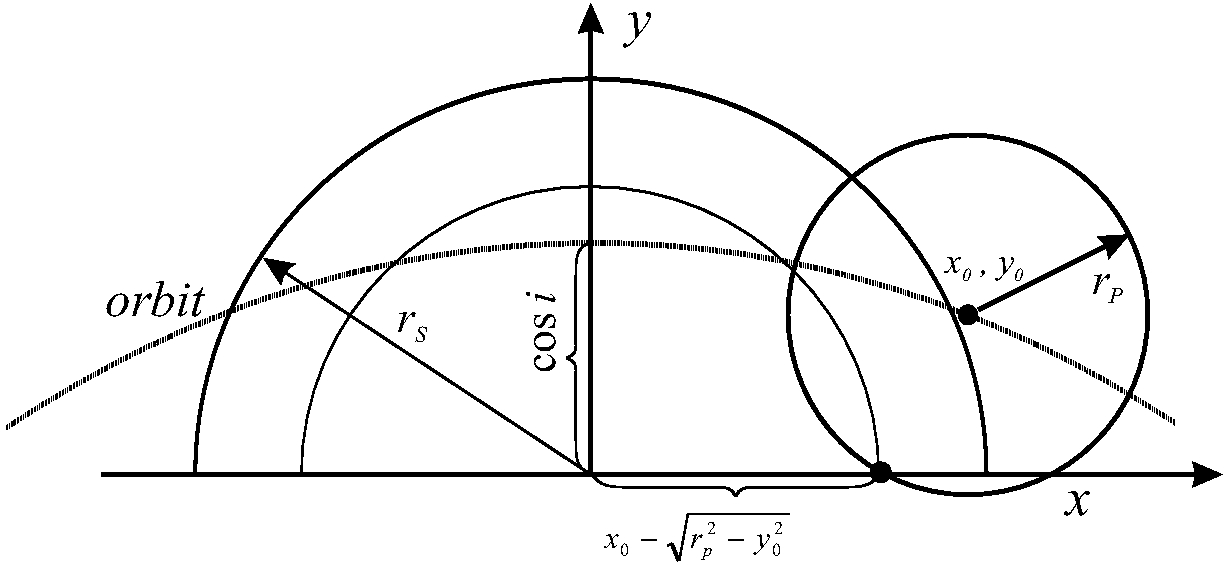}
 \caption{Geometry of partial transit in Case A.2}
 \label{fig02}
\end{figure}

\begin{figure}
 \centering
 \includegraphics[width=0.99\columnwidth]{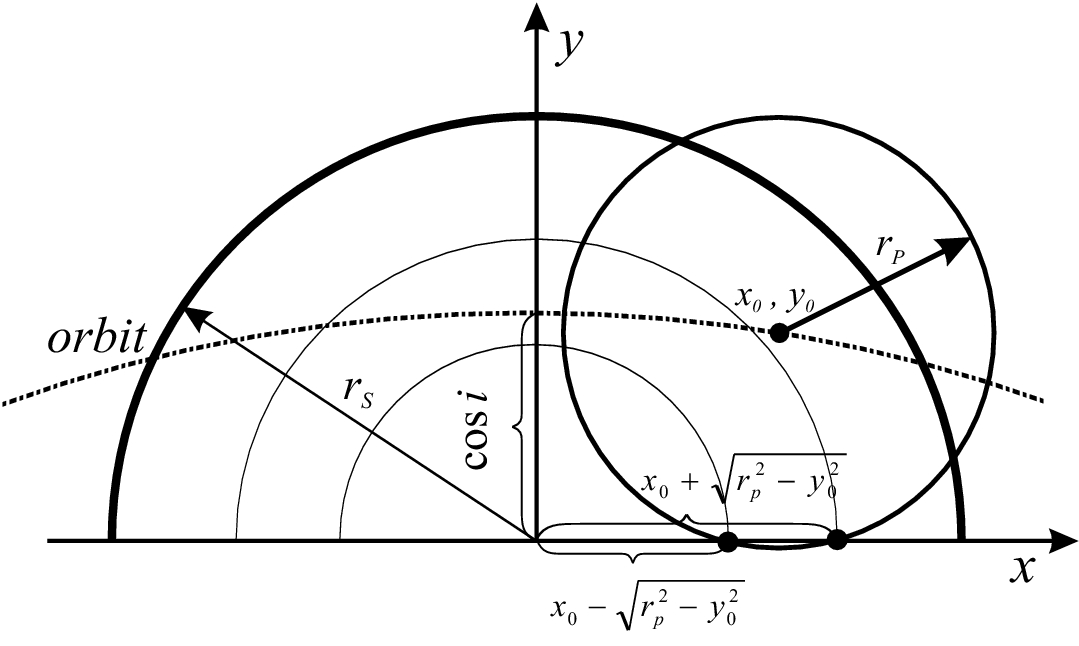}
 \caption{Geometry of partial transit in Case A.3}
 \label{fig03}
\end{figure}

\begin{description}
\item[(A.1)] If $x_{0} - \sqrt {r_{\rm{p}}^2 -y_{0}^2} \geq r_{\rm{p}}$ (Fig. \ref{fig01}) the light decreasing is calculated by (\ref{eq16}) 
where limits and expression for $2\gamma_{\rm{q}}(\varphi)$ are given in Table \ref{tab01} (Case A.1).

\item[(A.2)] If $x_{0} - \sqrt {r_{\rm{p}}^2 -y_{0}^2} \leq r_{\rm{p}} \leq x_{0} + \sqrt {r_{\rm{p}}^2 -y_{0}^2} $ then the integral $J(\varphi)$ 
in (\ref{eq16}) can be presented as a sum of two integrals $J_{1}(\varphi)+J_{2}(\varphi)$ (Fig. \ref{fig02}) whose limits and expressions for
2$\gamma_{\rm{q}}(\varphi)$ are given in Table \ref{tab01} (Case A.2).

\item[(A.3)] If $x_{0} + \sqrt {r_{\rm{p}}^2 -y_{0}^2} \leq r_{\rm{p}}$ the integral $J(\varphi)$ (Fig. \ref{fig03}) is a sum of three integrals 
$J_{1}(\varphi)+J_{2}(\varphi)+J_{3}(\varphi)$ whose limits and expressions for $2\gamma_{\rm{q}}(\varphi)$ are given in Table \ref{tab01} 
(Case A.3).
\end{description}

\begin{table*}
\caption{Integration limits and expressions for the integrands used for our direct problem solution} \label{tab01}
\begin{tabular}{ c c c c c}
\hline
Integrals& Cases & $q_{\rm{min}}(\varphi)$  & $q_{\rm{max}}(\varphi)$   & 2$\gamma_{\rm{q}}(\varphi)$  \\
\hline\hline
$J(\varphi)$     & A.1     & $b - r_{\rm{p}}$ & $r_{\rm{s}}$ & $\arccos {(x_{2}/q)} - \arccos{(x_{1}/q)}$ \\
$J(\varphi)$     & B.1     & $b - r_{\rm{p}}$ & $b + r_{\rm{p}}$ & $\arccos {(x_{2}/q)} - \arccos{(x_{1}/q)}$ \\
\hline
$J_{1}(\varphi)$ & A.2, A.3, B.2  & $b-r_{\rm{p}}$  &  $x_{0} - \sqrt {r_{\rm{p}}^2 -y_{0}^2} $    &  $ \arccos {(x_{2}/q)} - \arccos{(x_{1}/q)}$ \\
$J_{1}(\varphi)$ & C.1, C.2  &  0  &  $r_{p}-b$  &  $2\pi$ \\
\hline
$J_{2}(\varphi)$ & A.2  &$x_{0} - \sqrt {r_{\rm{p}}^2 -y_{0}^2} $    &    $r_{\rm{s}}$  &  $ \arccos {(x_{2}/q)} + \arccos{(x_{1}/q)}$ \\
$J_{2}(\varphi)$ & A.3, B.2  &$x_{0} - \sqrt {r_{\rm{p}}^2 -y_{0}^2} $ & $x_{0} + \sqrt {r_{\rm{p}}^2 -y_{0}^2} $ & $ \arccos {(x_{2}/q)} + \arccos{(x_{1}/q)}$ \\
$J_{2}(\varphi)$ & C.1  &$r_{p}-b$  &  $\sqrt {r_{p}^2 -y_{0}^2}-x_{0}$  & $ 2\pi - [\arccos {(|x_{1}|/q)} - \arccos{(|x_{2}|/q)}]$ \\
$J_{2}(\varphi)$ & C.2  &$r_{p}-b$    &    $\sqrt {r_{p}^2 -x_{0}^2}-y_{0}$  & $2\pi - [ \arccos {(|x_{1}|/q)} - \arccos{(|x_{2}|/q)}]$ \\
\hline
$J_{3}(\varphi)$ & A.3  &$x_{0} + \sqrt {r_{\rm{p}}^2 -y_{0}^2} $    &  $r_{\rm{s}}$  &  $ \arccos {(x_{2}/q)} - \arccos{(x_{1}/q)}$ \\
$J_{3}(\varphi)$ & B.2  &$x_{0} + \sqrt {r_{\rm{p}}^2 -y_{0}^2} $ &  $b+r_{\rm{p}}$  &  $ \arccos {(x_{2}/q)} - \arccos{(x_{1}/q)}$ \\
$J_{3}(\varphi)$ & C.1  &$\sqrt {r_{p}^2 -y_{0}^2}-x_{0}$  & $\sqrt {r_{p}^2 -x_{0}^2}-y_{0}$ & $ 2\pi - \arccos {(|x_{1}|/q)} - \arccos{(|x_{2}|/q)}$ \\
$J_{3}(\varphi)$ & C.2  &$\sqrt {r_{p}^2 -x_{0}^2}-y_{0}$  & $\sqrt {r_{p}^2-y_{0}^2}-x_{0}$ & $2\pi - \arcsin {(x_{1}/q)} - \arcsin{(|x_{2}|/q)}$\\
\hline
$J_{4}(\varphi)$ & C.1  &$\sqrt {r_{p}^2 -x_{0}^2}-y_{0}$  & $\sqrt {r_{p}^2 -x_{0}^2}+y_{0}$ & $ \arccos {(x_{2}/q)} + \arccos{(x_{1}/q)}$ \\
$J_{4}(\varphi)$ & C.2  &$\sqrt {r_{p}^2 -y_{0}^2}-x_{0}$  & $\sqrt {r_{p}^2 -y_{0}^2}+x_{0}$ & $ \arccos {(x_{2}/q)} + \arccos{(x_{1}/q)}$ \\
\hline
$J_{5}(\varphi)$ & C.1  &$\sqrt {r_{p}^2 -x_{0}^2}+y_{0}$  & $x_{0}+\sqrt {r_{p}^2-y_{0}^2} $ & $ \arccos {(x_{2}/q)} + \arccos{(x_{1}/q)}$ \\
$J_{5}(\varphi)$ & C.2  &$\sqrt {r_{p}^2 -y_{0}^2}+x_{0}$  & $y_{0}+\sqrt {r_{p}^2-x_{0}^2} $ & $ \arccos {(x_{2}/q)} - \arccos{(x_{1}/q)}$ \\
\hline
$J_{6}(\varphi)$ & C.1  &$x_{0} + \sqrt {r_{p}^2 -y_{0}^2} $  &  $b+r_{p}$  & $ \arccos {(x_{2}/q)} - \arccos{(x_{1}/q)}$ \\
$J_{6}(\varphi)$ & C.2  &$y_{0} + \sqrt {r_{p}^2 -x_{0}^2} $    &  $b+r_{p}$   & $ \arccos {(x_{2}/q)} - \arccos{(x_{1}/q)}$ \\
\hline
\end{tabular}
\end{table*}

\subsubsection{Case B: Total transits out the stellar center}

\begin{figure}
 \centering
 \includegraphics[width=0.99\columnwidth]{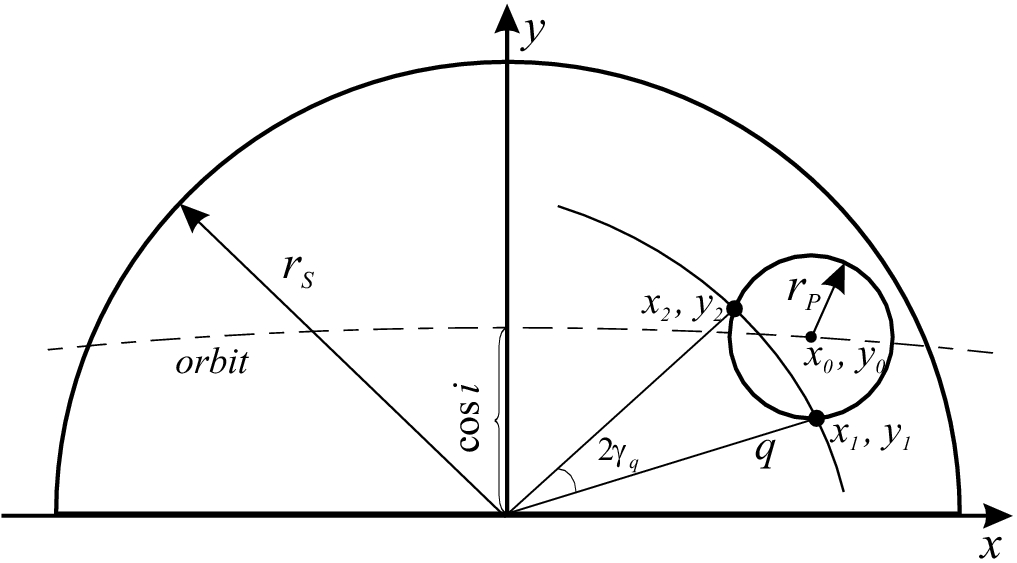}
 \caption{Geometry of the total transit, Case B}
 \label{fig04}
\end{figure}

If the orbital inclination is into the range $i_{2}~\leq~i~\leq~i_{3}$ where

\begin{equation}\label{eq24}
i_{3} = \arccos {(r_{\rm{p}})}
\end{equation}

\noindent then the transit develops from partial to total. In this case the planet does not cover the star center at any phase (Fig. \ref{fig04}).

The phases of the inner contact star-planet are $-\varphi_{2}$ (during planet' entering) and $+\varphi_{2}$ (during planet' exit) where

\begin{equation}\label{eq25}
\varphi_{2}= \frac {1}{2 \pi} \arcsin
\frac{\sqrt{(r_{\rm{s}}-r_{\rm{p}})^2- \cos i^2} } {\sin i} .
\end{equation}

Into the phase range [$\varphi_{2}, \varphi_{1}$] the transit is partial and geometry is similar to the Case A. Into the phase range 
[0,~$\varphi_{2}$] the transit is total.

\begin{description}
\item[(B.1)] If $y_{0} > r_{\rm{p}}$ then the integral $J(\varphi)$ is calculated by (\ref{eq16}) where limits and expression for
$2\gamma_{\rm{q}}(\varphi)$ are given in Table \ref{tab01} (Case B.1).

\item[(B.2)] If $y_{0} \leq r_{\rm{p}}$ the integral $J(\varphi)$ is a sum of three integrals 
$J_{1}(\varphi)+J_{2}(\varphi)+J_{3}(\varphi)$ whose attributes are given in Table \ref{tab01} (Case B.2).
\end{description}

\begin{figure}
 \centering
 \includegraphics[width=0.99\columnwidth]{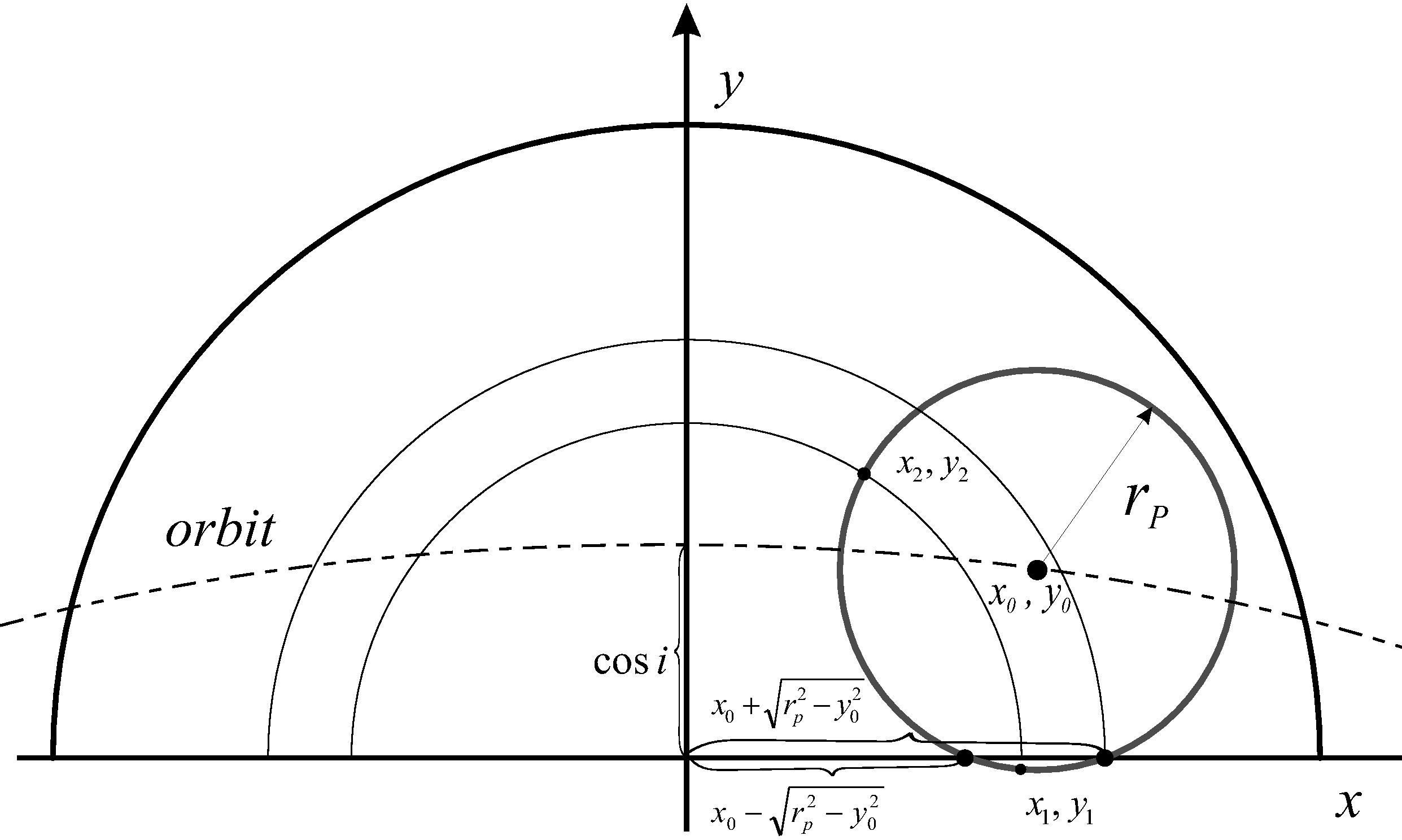}
 \caption{Geometry of the total transit for $y_{0}~\leq~r_{\rm{p}}$, Case B.2}
 \label{fig05}
\end{figure}

\subsubsection{Case C: Total transit through the stellar center}

If the orbital inclination is into the range $i_{3} \leq i \leq 90\degr$ the transit develops from partial to total and finally the planet covers 
the stellar center.

The phases at which the planet limb touches the stellar center are $-\varphi_{3}$ (before the transit center) and $+\varphi_{3}$ (after the transit 
center) where

\begin{equation}\label{eq27}
\varphi_{3}= \frac {1}{2 \pi} \arcsin \frac { \sqrt{r_{\rm{p}}^2-\cos i^2} } { \sin i}  .
\end{equation}

Into the phase range [$\varphi_{2}, \varphi_{1}$] the transit is partial and the geometry is similar to the Case A.

If $y_{0} \geq r_{\rm{p}}$ then the integral $J(\varphi)$ is calculated by (\ref{eq16}) where limits and expression for $2\gamma_{\rm{q}}(\varphi)$ 
are given in Table \ref{tab01} (Case A.1).

If $y_{0} \leq r_{\rm{p}}$ and $x_{0} + \sqrt {r_{\rm{p}}^2 -y_{0}^2} \geq r_{\rm{s}}$ then the integral $J(\varphi)$ is presented as a sum of two 
integrals $J_{1}(\varphi)+J_{2}(\varphi)$ whose attributes are the same as those of the Case A.2 (Table \ref{tab01}).

If $y_{0} \leq r_{\rm{p}}$ and $x_{0} + \sqrt {r_{\rm{p}}^2 -y_{0}^2} \leq r_{\rm{s}}$ then the integral $J(\varphi)$ is presented as a sum of three 
integrals $J_{1}(\varphi)+J_{2}(\varphi)+J_{3}(\varphi)$ whose attributes are the same as those of the Case A.3 (Table \ref{tab01}).

Into the phase range [$\varphi_{3}, \varphi_{2}$] the transit is total, outside the stellar center. The geometry is similar to Case B.

If $y_{0} > r_{\rm{p}}$ then the integral $J(\varphi)$ is calculated by (\ref{eq16}) where limits and expression for $2\gamma_{\rm{q}}(\varphi)$ 
are given in Table \ref{tab01} (Case B.1).

If $y_{0} \leq r_{\rm{p}}$ then the integral $J(\varphi)$ is presented as a sum of three integrals $J_{1}(\varphi)+J_{2}(\varphi)+J_{3}(\varphi)$ 
whose attributes are the same as those of the Case B.2  (Table \ref{tab01}).

Into the phase range [0, $\varphi_{3}$] the planet covers the stellar center and there are three subcases depending on the phase

\begin{equation}\label{eq28}
\varphi_{4}= \frac {1}{2 \pi} \arctan ( \cos i)
\end{equation}

\noindent that corresponds to the moment when $x_{0} = y_{0}$.

\begin{figure}
 \centering
 \includegraphics[width=0.99\columnwidth]{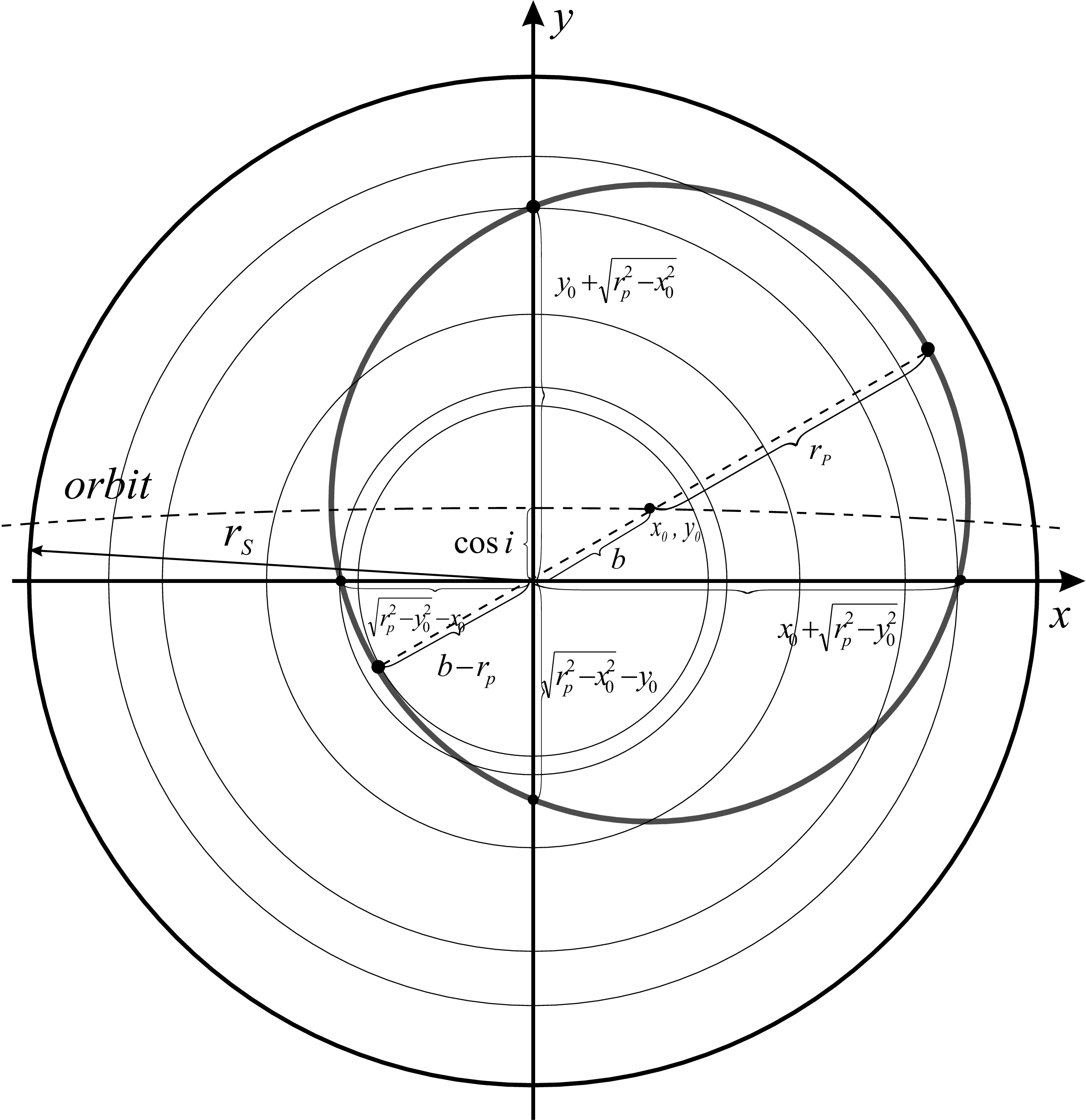}
 \caption{Case C.1 and C.3, total transit through stellar center and $x_{0}~\geq~y_{0}$}
 \label{fig06}
\end{figure}

\begin{figure}
 \centering
 \includegraphics[width=0.99\columnwidth]{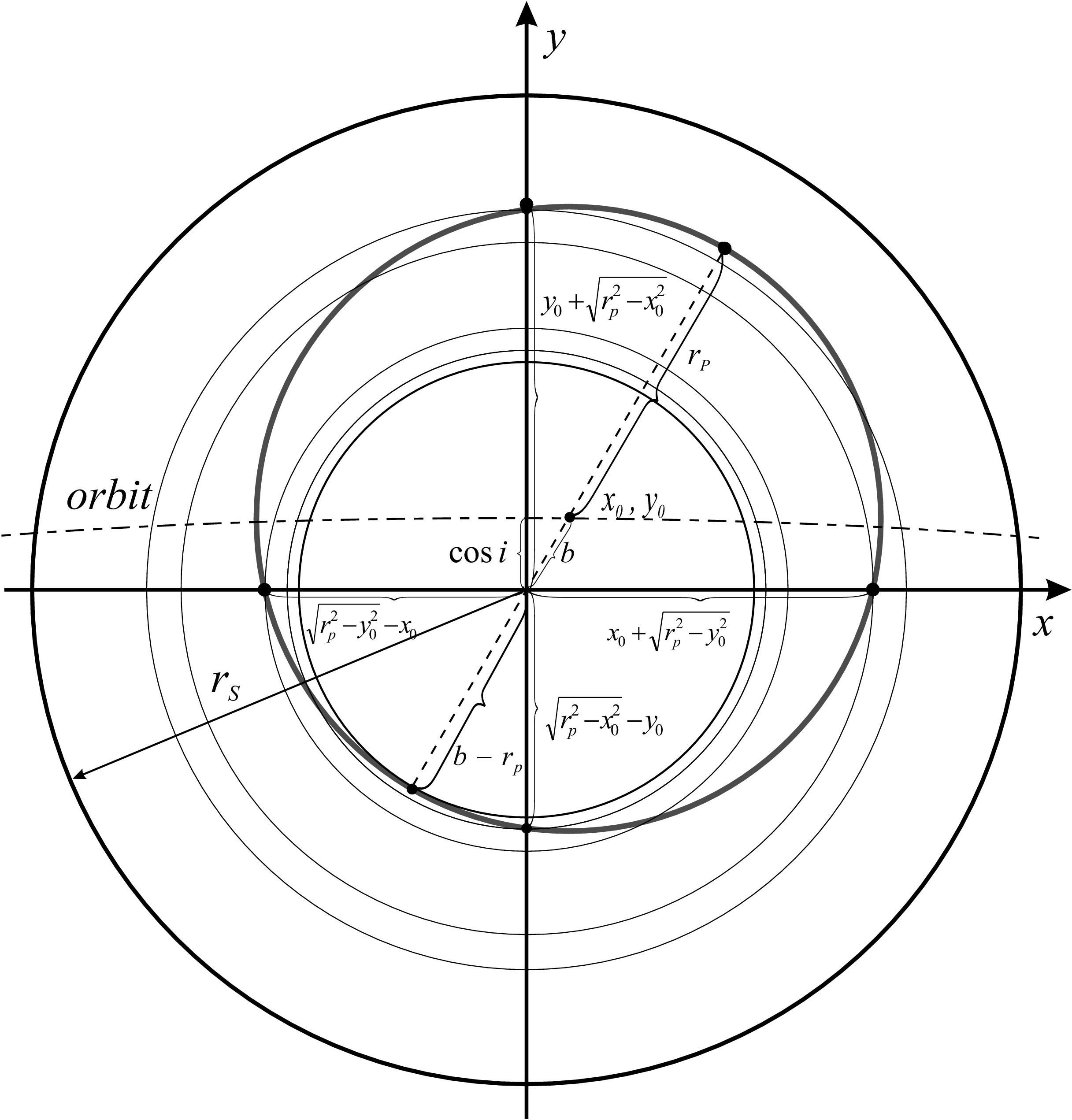}
 \caption{Case C.2, total transit through stellar center and $x_{0}~\leq~y_{0}$}
 \label{fig07}
\end{figure}

\begin{description}
\item[(C.1)] If $\varphi_{4} \leq \varphi_{3}$ then into the phase range [$\varphi_{4}, \varphi_{3}$] (for which $x_{0} \geq y_{0}$) the integral 
$J(\varphi)$ is presented as a sum of six integrals $J_{1}(\varphi)+J_{2}(\varphi)+J_{3}(\varphi)+J_{4}(\varphi)+J_{5}(\varphi)+J_{6}(\varphi)$
(Fig. \ref{fig06}) whose attributes are given in Table \ref{tab01} (Case C.1).

\item[(C.2)] If $\varphi_{4} \leq \varphi_{3}$ then into the phase range [0, $\varphi_{4}$] (for which $x_{0} \leq y_{0}$) the integral 
$J(\varphi)$ is presented as a sum of six integrals $J_{1}(\varphi)+J_{2}(\varphi)+J_{3}(\varphi)+J_{4}(\varphi)+J_{5}(\varphi)+J_{6}(\varphi)$
(Fig. \ref{fig07}) whose attributes are given in Table \ref{tab01} (Case C.2).

\item[(C.3)] If $\varphi_{4} \geq \varphi_{3}$ then into the whole phase range [0, $\varphi_{3}$] is fulfilled $x_{0} \geq y_{0}$ and the integral 
$J(\varphi)$ is presented as a sum of six integrals $J_{1}(\varphi)+J_{2}(\varphi)+J_{3}(\varphi)+J_{4}(\varphi)+J_{5}(\varphi)+J_{6}(\varphi)$
(Fig. \ref{fig06}) whose attributes are the same as those of Case C.1.
\end{description}

Note: The condition $\varphi_{4} \geq \varphi_{3}$ is satisfied for $i_{3} \leq i \leq i_{4}$ where

\begin{equation}\label{eq29}
i_{4} = \arccos {\frac {r_{\rm{p}}}{\sqrt{2-r_{\rm{p}}^{2}}}} .
\end{equation}

Finally, it should be noted that for the arbitrary limb-darkening law (\ref{eq06}) the stellar luminosity L$_{\rm{s}}$ is given by the integral

\begin{equation}\label{eq30}
L_{\rm{s}}=\int \limits_{0}^{R_{\rm{s}}} I_{0} f(u_{\rm{j}}, \mu) 2 \gamma_{\rm{r}} r dr  .
\end{equation}

\noindent It has analytical solution for all known limb-darkening functions. Particularly, for the wide-used limb-darkening laws (equations from 
(\ref{eq01}) to (\ref{eq05})) the stellar luminosity is calculated by the formulae

\begin{eqnarray}\label{eq31}
L_{\rm{s}} & = & \pi R_{\rm{s}}^2 I_{0} \left( 1-\frac{u}{3} \right) \qquad\qquad\qquad \mbox{linear~law} \\
L_{\rm{s}} & = & \pi R_{\rm{s}}^2 I_{0} \left( 1-\frac{u_1}{3} - \frac{u_2}{6} \right) \qquad\quad \mbox{quadratic~law} \nonumber \\
L_{\rm{s}} & = & \pi R_{\rm{s}}^2 I_{0} \left( 1-\frac{u_1}{3} - \frac{u_2}{5} \right) \qquad\quad \mbox{squared-root~law} \nonumber \\
L_{\rm{s}} & = & \pi R_{\rm{s}}^2 I_{0} \left( 1-\frac{u_1}{3} + 2\frac{u_2}{9} \right) \qquad\;\, \mbox{logarithmic~law} \nonumber \\
L_{\rm{s}} & = & \pi R_{\rm{s}}^2 I_{0} \left( 1-\frac{u_1}{5} - \frac{u_2}{3} - 3 \frac{u_3}{7} - \frac{u_4}{2}\right) \mbox{``nonlinear''~law} \nonumber
\end{eqnarray}

\section{The code \textsc{TAC-maker}}

The integral in equation (\ref{eq16}) cannot be solved analytically. That is why we had to carry out a numerical solution of this integral.

For this reason we wrote the code \textsc{TAC-maker} (\textbf{T}ransit \textbf{A}nalytical \textbf{C}urve) whose input parameters are:
\begin{itemize}

\item radius of the orbit $a$;
\item Period $P$ and initial epoch $T_0$;
\item radius of the star $R_{\rm{s}}$;
\item radius of the planet $R_{\rm{p}}$;
\item orbital inclination $i$;
\item temperature of the star $T_{\rm{s}}$;
\item temperature of the planet $T_{\rm{p}}$;
\item coefficients of the limb-darkening $u_{\rm{j}}$;
\item step in phase $PH_{\rm{step}}$;
\item parameter of precision $PP_{\rm{TAC}}$ of the numerical calculations of the integrals.
\end{itemize}

The code \textsc{TAC-maker} provides the possibility to choose the limb-darkening law from a list of the known wide-spread functions (1 -- 5), or 
to write arbitrary function $f(u_{\rm{j}}, \mu$). This is a significant advantage of the proposed approach as even now the accuracy of the most
used quadratic limb-darkening law is worse than the achieved from \textit{Kepler} for large planets with $R_{\rm{p}} > 0.04 R_{\rm{s}}$ 
\citep{eastman13}.

Moreover, the code \textsc{TAC-maker} allows to obtain the stellar limb-darkening coefficients from the transit solution and to compare them with 
the theoretical values of \citet{claret04} calculated for different temperatures, surface gravities, metal abundances and micro-turbulence 
velocities.

The code is written in \textsc{Python~2.7} language with a Graphical User Interface. At the very beginning the code checks which condition 
(Case A, Case B, or Case C) is satisfied for the given combination of configuration parameters (stellar and planet radii and orbital inclination). 
After that the code calculates the characteristic phases $\varphi_{\rm{i}}$ for the respective case. Further the code makes numerical calculation 
of the integral (\ref{eq16}) for the current phase and chosen function $f(u_{\rm{j}}, \mu)$ using the \textsc{scipy} package. Finally, the code 
repeats the procedure for each phase of the corresponding phase ranges. The output results flow as data file (phase $\varphi$, flux $F$).

The code allows to search for solutions of observed transits by the method of trials and errors varying the input parameters. The data
file might be in format magnitude or flux and $HJD$ or phase. An estimate of the fit quality is the calculated value of $\chi^2$. Moreover,
the two plots of the current solution showing the observational data with the synthetic transit as well as the corresponding phase distribution
of the residuals allow fast finding of a good fit.

\subsection{Validation of \textsc{TAC-maker}}

To validate our approach we used comparison with the wide-spread \citetalias{mandel02} solution, particularly we compared the synthetic light curves 
generated by the code \textsc{TAC-maker} and those produced by the code \textsc{occultnl} for the same configuration parameters and the same 
limb-darkening law. For this purpose we applied the freely available version of the last code without any changes, particularly with its default 
precision, while our code worked with the default precision of the \textsc{scipy} package for the numerical calculations of the integrals.

\subsubsection{Comparison for linear limb-darkening law}

\begin{figure}
 \centering
 \includegraphics[width=0.99\columnwidth]{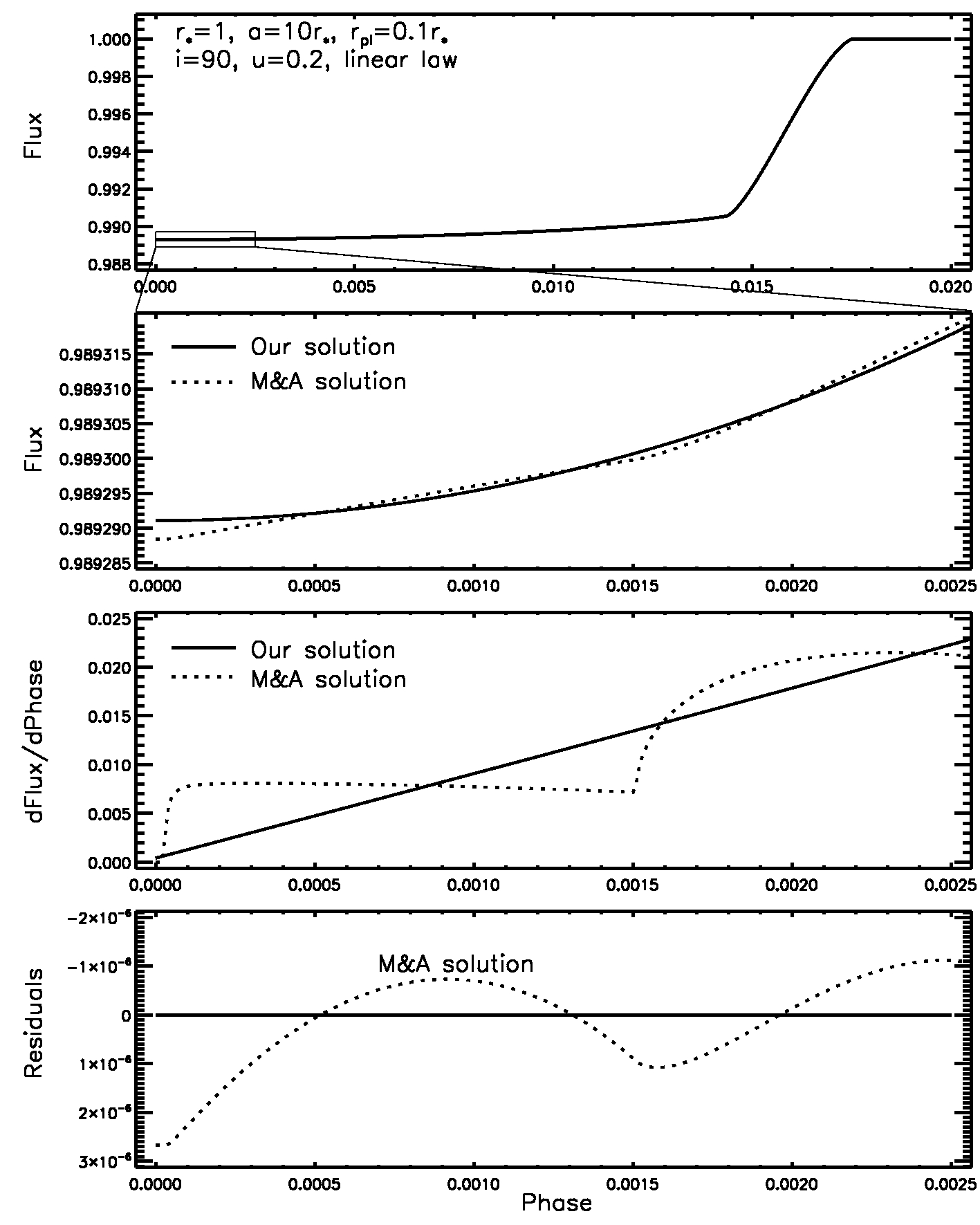}
 \caption{Comparison of the new solution with that of \citet{mandel02} for the same parameters: $i=90\degr$, $a=10$, $R_{\rm{s}}=1$, 
$R_{\rm{p}}=0.1$, $T_{\rm{s}}=5000$ K, $T_{\rm{p}}=0$. \textit{Top panel}: The coincidence of the two solutions for the linear limb-darkening law 
and moderate precision of the \citetalias{mandel02} method; \textit{Second panel}: The difference of the two solutions for small part of the 
transit in big scale; \textit{Third panel}: The first derivatives of the two solutions for small part of the transit in big scale; 
\textit{Bottom panel}: The residuals for the chosen part of the transit in big scale.}
 \label{fig08}
\end{figure}

We established that for linear limb-darkening law the two synthetic transits coincide (Fig. \ref{fig08}, top). However, the detailed review reveals 
the meandering course of the \citetalias{mandel02} solution around the smooth course of the \textsc{TAC-maker} transit curve (Fig. \ref{fig08}, 
second panel). The phase derivatives of the fluxes (Fig. \ref{fig08}, third panel) exhibit more clearly which one of the two solutions is precise: 
the derivative of the \textsc{TAC-maker} solution has almost linear course while that of the \citetalias{mandel02} solution reveals plantigrade 
shape. This result allows us to assume that our solution of the direct problem for the planet transit in the case of linear limb-darkening law 
is more accurate than that of \citetalias{mandel02}.

For more detailed comparison of the two approaches we analyzed the differences (residuals) between the flux values of the \textsc{TAC-maker} 
solution and the \citetalias{mandel02} solution for the same configuration parameters and phases. We established that they depend on different 
parameters (limb-darkening coefficients, inclination, phase, planet radius, planet temperature).

For linear limb-darkening law the residuals vary with the phase by oscillating way around level 0 (Fig. \ref{fig09}) which analysis led us to the 
following conclusions.

\begin{description}

\item[(a)] The frequencies and amplitudes of the oscillating residuals depend on the limb-darkening coefficients (Fig. \ref{fig09}, top). 
The residuals are zero for $u=0.0$ that means that the \citetalias{mandel02} solution is precise for uniform star.

\begin{figure}
 \centering
 \includegraphics[width=0.99\columnwidth]{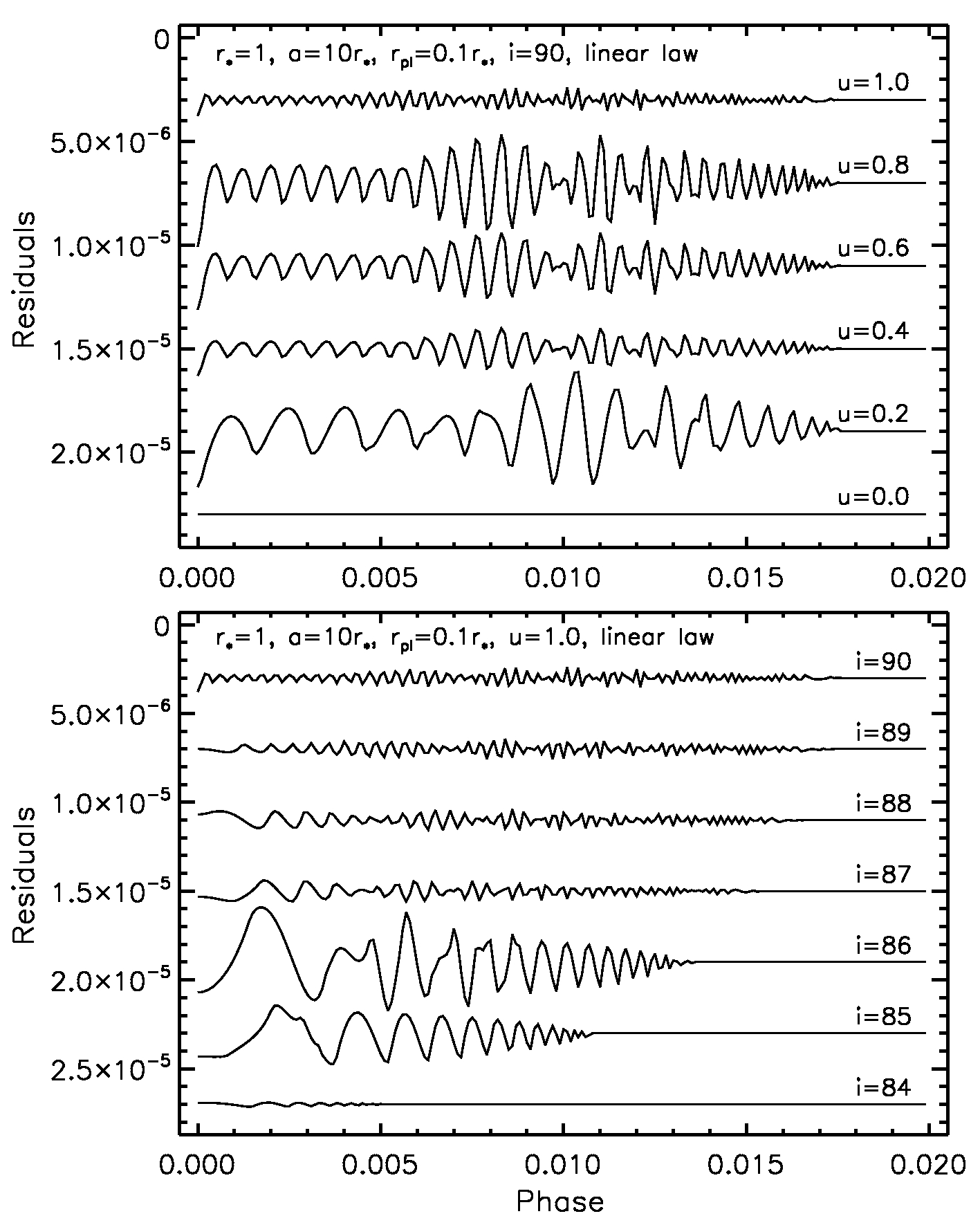}
 \caption{Residual oscillations in the case of linear limb-darkening law: \textit{top} for $i=90\degr$ and different linear limb-darkening 
coefficients; \textit{bottom} for different orbital inclinations and $u=1.0$ (all residuals oscillate around level 0 but are shifted vertically 
for a good visibility).}
 \label{fig09}
\end{figure}

\item[(b)] The frequencies of the residual oscillations decrease to the central part of the transit for all orbital inclinations, especially for 
small inclinations (Fig. \ref{fig09}, bottom).

\item[(c)] The amplitudes of the residuals are bigger for partial transits than for total ones (Fig. \ref{fig09}, bottom).

\item[(d)] The residual values increase with the planet radius (as it should be expected).

\end{description}

\subsubsection{Comparison for nonlinear limb-darkening laws}

We present comparison with those nonlinear limb-darkening laws which are available in the code \textsc{occultnl}.

\begin{figure}
 \centering
 \includegraphics[width=0.99\columnwidth]{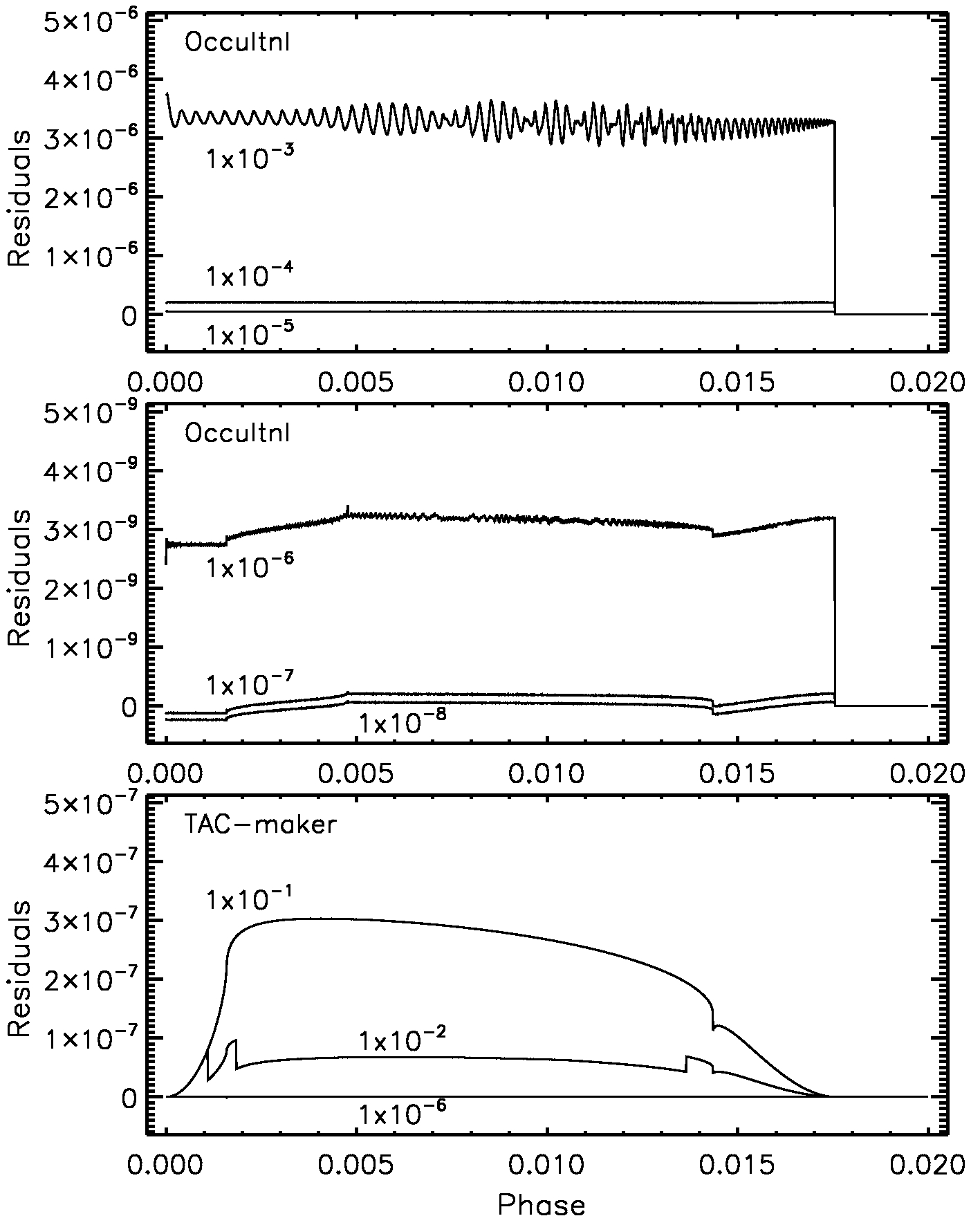}
 \caption{Quality of the synthetic transit for different values of the precision parameters for the case of quadratic limb darkening
law and equal coefficients. \textit{Top} and \textit{middle} panels: residuals for the code \textsc{occultnl} corresponding to different values
of $PP$; \textit{bottom} panel: residuals for the code \textsc{TAC-maker} corresponding to different values of $PP_{\rm{TAC}}$.}
 \label{fig10}
\end{figure}

\begin{description}

\item[(1)] The comparison of the synthetic light curves generated for the same configuration parameters and quadratic limb-darkening law revealed 
that those produced by the code \textsc{occultnl} exhibited (Fig. \ref{fig10}): (i) small-amplitude oscillations similar to those for the linear 
limb-darkening law; (ii) the fluxes into the transit are systematically smaller than ours (the underestimation increases with the coefficient $u_2$ 
of the quadratic term).

\item[(2)] The comparison of the synthetic light curves generated for the same configuration parameters and squared root limb-darkening law 
revealed that those produced by the code \textsc{occultnl} suffered from the same disadvantages as those for quadratic limb-darkening law 
(Fig. \ref{fig10}).

\end{description}

The detailed comparison (validation) of our approach with the wide-spread \citetalias{mandel02} method demonstrated that the difference between 
them did not exceed $10^{-6}$ for the considered three types limb-darkening laws. This leads to the conclusion that for variety of model parameters 
the new approach gives reasonable and expected synthetic transits which practically coincide with those of the \citetalias{mandel02} solution.

\subsection{Precision vs computational speed}

There is a possibility to increase the precision of the synthetic transits generated by the code \textsc{occultnl} (E.~Agol, private communication). 
Its convergence criterion is the maximum change to be less than the product of two multipliers, the factor $PP=10^{-n}$ ($n$ is an integer) and 
the transit depth.

We established that the decreasing of $PP$ from its default value $10^{-3}$ leads to the following effects (Fig. \ref{fig10}, top and middle panels):
 (i) decreasing of the amplitudes of the oscillating residuals; (ii) increasing of their frequencies; (iii) decreasing of the underestimation of the 
fluxes for the whole transit.

On the other hand the default value of $PP_{\rm{TAC}}$ is $1.4\times10^{-8}$ (as it is defined for the package \textsc{scipy}). Its increasing 
leads to some small imperfections of the numerical calculations made by our code (Fig. \ref{fig10}, bottom).

The two considered approaches of the direct problem solution lead to different definitions of the precision parameters of the corresponding codes. 
Therefore, it is not reasonable to compare the precision of the generated synthetic transits for equal values of the parameters $PP_{\rm{TAC}}$ 
and $PP$. Though, we established that the two methods allow one to reach the desired (practically arbitrary) precision of calculations by 
appropriate choice of the corresponding precision parameter. Moreover, for the best precisions, the differences between the transit fluxes 
generated with the two codes are below $10^{-10}$.

The computational speed is another important characteristic of the codes for transit synthesis, especially when they are used for the solution of 
the corresponding inverse problem. In principle, this speed depends on two parameters: time precision (step in phase) and space precision 
(size of the differential element of the integrals).

The thorough comparison of the computational speeds of different codes for the solution of the same problem requires these codes to be written in 
the same languages, to be run on the same computers and for the equal model parameters. That is why it is not reasonable to compare the 
computational speed of our code with the others.

Nevertheless, we made some tests which revealed the following results.

\begin{description}
\item[(a)] The reducing of $PP$ (i.e. the increasing of the precision of the synthetic transit) by one order leads to 3-5 times increase of the 
computational time while the reducing of $PP$ from $10^{-3}$ to $10^{-8}$ (with five orders) requires around 270 times longer computational time 
for the code \textsc{occultnl}. 
\item[(b)] The decreasing of $PP_{\rm{TAC}}$ from $10^{-1}$ to $10^{-8}$ (with seven orders) requires around 15 times longer computational time. 
\item[(c)] The formal comparison of the absolute computational speeds of the \textsc{Python} code \textsc{TAC-maker} and \textsc{IDL} code 
\textsc{occultnl} revealed that our code is slightly faster for high-precision calculations while for the low-precision calculations the code
\textsc{occultnl} is faster than the code \textsc{TAC-maker}. From these results as well as from the consideration that each \textsc{IDL} code 
is faster than its \textsc{Python} version we conclude that the code \textsc{TAC-maker} possesses good computational speed for the high-precision 
calculations. 
\item[(d)] The new subroutine \textsc{exofast\_occultquad} \citep{eastman13} is around 2 orders faster than its progenitor \textsc{occultquad}. 
Moreover, we established that \textsc{exofast\_occultquad} (with \textsc{IDL, Fortran} and \textsc{Python} versions) is considerably improved
version of \textsc{occultquad} not only concerning the computational speed but also concerning the precision (we had found some bugs of 
\textsc{occultquad}). Although the computational speed of \textsc{exofast\_occultquad} is higher than those of \textsc{TAC-maker} and 
\textsc{occultnl}, it should be remembered that \textsc{exofast\_occultquad} can produce transits only for quadratic limb-darkening law.
\end{description}

\subsection{A possibility to fit the planet temperature}

The previous transit solutions consider the planet as a black lens and the corresponding codes do not fit the planet temperature $T_{\rm{p}}$. 
As a result only a formal parameter, the equilibrium temperature of the planet depending on the stellar heating (i.e., on stellar temperature and 
planet distance), can be calculated (out of the procedure of transit modelling).

The increasing precision of the observations raises the problem of the planet temperature, i.e. to study the effect of the planet temperature on 
the transit and to search for a possibility to determine this physical parameter from the observations.

The new approach allowed such tests to be directly carried out because $T_{\rm{p}}$ is an input parameter of its own (unlike the previous methods). 
As a result, we established that increasing of the planet temperature from 0 K to 1000 K causes the decreasing of the transit depth up to $10^{-6}$ 
while the increasing of the planet temperature from 0 K to 2000 K leads to shallower transit up to $10^{-5}$. The contribution of the planet 
temperature on the transit depth increases both with the ratio $R_{\rm{p}}/R_{\rm{s}}$ and limb-darkening coefficients. Moreover, it rapidly 
increases with wavelength.

The obtained estimations revealed that the effect of the planet temperature is lower than the precision of the present optical photometric 
observations, even than that of the \textit{Kepler} mission. Hence, the determination of the real planet temperature from the observed transit in 
optics is postponed for the near future.

However, the sensibility of the observations at longer wavelengths to the planet emission is higher. Recently, there have been reports that the 
observed IR fluxes of some hot Jupiter systems during occultation are higher than those corresponding to their equilibrium temperatures 
\citep{gill09, gibs10, croll10a}. These results imply that the determination of the planet temperature from the observed transits is forthcoming. 
Until such observational precision is reached the code \textsc{TAC-maker} could be used for theoretical investigations of the effects of different 
thermal processes (as Ohmic heating, tidal heating, internal energy sources, etc.) on the planet
transits.

\section{Conclusions}

Although the analytical solutions of the problems are very useful, the modern astrophysical objects and configurations are quite complex to allow 
analytical descriptions. Instead of analytical solutions we are now able to use fast numerical computations.

This paper presents a new solution of the direct problem of the transiting planets. It is based on the transformation of the double integrals 
describing the light decrease during the transit to linear ones. We created the code \textsc{TAC-maker} for generation of synthetic transits
by numerical calculations of the linear integrals.

The validation of our approach was made by comparison with the results of the wide-spread \citetalias{mandel02} method for modelling of planet 
transits. It was demonstrated that our method gave reasonable and expected synthetic transits for linear, quadratic and squared-root limb-darkening 
laws and arbitrary combinations of input parameters.

The main advantages of our approach for the planet transits are:

\begin{description}
\item[(1)] It gives a possibility, for the first time, to use an arbitrary limb-darkening law $f(u_{\rm{j}}, \mu$) of the host star; 
\item[(2)] It allows acquisition of the stellar limb-darkening coefficients from the transit solution and comparison with the theoretical values of
\citet{claret04}; 
\item[(3)] It gives a possibility, in principle, to determine the planet temperature from the observed transits. Our estimations reveal that the 
effect of the non-zero planet temperature to the transit depth is lower than the precision of the present optical photometric observations.
However, the higher sensibility of the observations at longer wavelengths (IR) to the planet emission implies that the determination of the planet 
temperature from the observed transits is forthcoming.
\end{description}

These properties of our approach and the practically arbitrary precision of the calculations of the code \textsc{TAC-maker} reveal that our 
solution of the planet transit problem is able to meet the challenges of the continuously increasing photometric precision of the ground-based and 
space observations.

We plan to build an inverse problem solution for the planet transit (for determination of the configuration parameters) on the basis of our direct 
problem solution and by using the derived simple analytical expressions in this paper to obtain initial values of the fitted parameters.

The code \textsc{TAC-maker} is available for free download from the Astrophysics Source Code Library\footnote{http://asterisk.apod.com/wp/} 
or its own site\footnote{http://astro.shu-bg.net/software/TAC-maker}.

\section*{Acknowledgments}

The research was supported partly by funds of projects DO~02-362, DO~02-85, and DDVU 02/40-2010 of the Bulgarian Scientific Foundation.
We are very grateful to Eric Agol, the referee of the manuscript, for the valuable recommendations and useful notes.

\label{lastpage}

\end{document}